\documentclass[preprint,preprintnumbers,amsmath,amssymb]{revtex4}
\usepackage{graphicx}
\usepackage{dcolumn}
\usepackage{bm}
\date{today}
\begin{document}
\title{ A note on constant curvature solutions in cylindrically symmetric metric $f(R)$ Gravity  }
\author{D.Momeni\footnote{Corresponding author}}
\email{dmomeni@phymail.ut.ac.ir} \affiliation{ Department of
Physics,Faculty of science,Islamic Azad University, Karaj
Branch.Iran, Karaj, Rajaei shahr, P.O.Box: 31485-313 }
\author{H. Gholizade}
\email{gholizade@ut.ac.ir} \affiliation{ Department of Engineering
Science, Faculty of Engineering, University of Tehran, Tehran, PO
Box 11155-4563, Iran }

\begin{abstract}
   In  the previous work  we  introduced a new  static cylindrically symmetric vacuum solutions in Weyl
coordinates in the context of the metric f(R) theories of
gravity\cite{1}. Now we  obtain a  2-parameter family of exact
solutions which contains cosmological constant and a new parameter
as $\beta$. This solution corresponds to a constant  Ricci scalar.
We proved that in $f(R)$  gravity , the constant curvature solution
in cylindrically symmetric cases is only one member of the most
generalized Tian family in GR. We show that our constant curvature
exact solution is applicable to the exterior of a string.
Sensibility of stability under initial conditions is discussed.

\end{abstract}
 \maketitle
\section{\label{sec:level1}Introduction}
Any phenomenal problem in theoretical physics commitment using of
any thinkable method to attack. Acceleration of universe in now era
is the most challenge for all branch of physics. If we are
interested only on attribute it to a matter field, we envisage with
different kind of models which we called them physical models or
dark fluids.We present one of them as a simple phase transition in a
known metric in [9]. Another approach is encroach to geometrical
parts of Einstein field equations and working with a most general
action. all these different proposals have been put forward with the
hope of obtaining a consistent theoretical background to the recent
observation of an expanding universe which is seemingly not
accessible through the Standard Model of cosmology.
 Modified or alternative theories of gravity is the paradigm under which all those theories which
differ from the Einsteinian gravity are studied. One of these
non-Einsteinian theories, that arose a lot of enthusiasm recently,
is the so-called f (R) gravity in which a function, f (R), replaces
the Einstein–Hilbert (gravitational) Lagrangian R. Later Buchdahl
published a paper about some Mathematical solutions with this form
of action[2]. These theories could be thought of as a special kind
of higher derivative gravitational theories. As in the case of the
Einstein–Hilbert action one could derive field equations in f (R)
gravity in two different approaches, the so-called metric and
Palatini approaches. But in f (R) actions, unlike the
Einstein–Hilbert action or its modified version (one with a
cosmological constant term), the field equations obtained by the two
approaches are not the same in general. In what follows we will be
interested only in metric f (R) theories of gravity in which
connection is dependent on metric $g_{\mu\nu}$ with respect to
which, the action is varied. In ordinary GR there are not that many
exact solutions of the field equations for a given symmetry. Being
higher derivative theory it is not unexpected to find more exact
solutions in f (R) gravity and this turns out to be the case for
spherically symmetric solutions [3].Constant curvature solutions in
 $f(R)$  gravity was studied by  Odintsov and collaborators
\cite{11}.Also there is a significant review of f(R) and other
modified gravities \cite{12} and viable models \cite{13} in
literatures . Former Looking for solutions with a different
symmetry, we obtained static cylindrically symmetric vacuum
solutions of the f (R) modified Einstein equations [1]. It is shown
how one can reduce the set of equations into a single equation which
could then be utilized to construct explicit solutions. Now we
generalized field equations with the same symmetry For constant
curvature solutions, using the general form of a cylindrically
symmetric solution in Weyl coordinates we find, among possible
solutions, a generalized form of zero curvature spacetime as well as
one new (non-zero curvature) solutions with one of their parameters
chosen so that it is related naively to the cosmological constant.
The stability of a non linear system of differential equations is a
significance problem. The modern method for study the stability
problem is theory of dynamical systems that first is used by Coley
in cosmological models \cite{14}.Folowing Coley we applied this
method for our model and obtained a constraint on the initial
condition which under it , the system of non linear equations are
stable. Our motivation in this paper is that we solved the non
linear equations and obtained Tian solution and  we prove that our
exact solution is applicable to the exterior of a string
\cite{5}.All mathematical calculations is done in detail .Therefore
this work complete our former paper and although contains some new
features of constant curvature solutions in $f(R)$ metric gravity in
cylindrical symmetry.Unlike our previous article we do not limit
ourselves to a specific gauge , rather we work in a general
framework. Finally we present a prediction for the general form of
this cylindrically symmetric solution in other modified $f(G)$
,$f(R,G)$ models of gravity.
\section{Field equations in $f$(R) gravity}
In this section we give a brief review of the field equations in
$f(R)$ gravity. The action for $f(R)$ gravity is given by
\begin{eqnarray}
S=\int(f(R)+\mathfrak{L}_{m})\sqrt{-g}d^{4}x
\end{eqnarray}
The field equation resulting from this action in the metric
approach, i.e. assuming that the connection is the Levi-Civita
connection and the
 variation is done with respect to the metric $g_{\mu\nu}$, is given by
\begin{eqnarray}
{G}_{\mu\nu}\equiv
{R}_{\mu\nu}-\frac{1}{2}{R}g_{\mu\nu}=\mathrm{T}^{g}_{\mu\nu}
+\frac{\mathrm{T}^{m}_{\mu\nu}}{F(R)}
\end{eqnarray}
where the gravitational stress-energy tensor is
\begin{eqnarray}
\mathrm{T}^{g}_{\mu\nu}=\frac{1}{F(R)}(\frac{1}{2}g_{\mu\nu}(f(R)-R
F(R))+F(R)^{;\alpha\beta}(g_{\alpha\mu}g_{\beta\nu}-g_{\mu\nu}g_{\alpha\beta}))
\end{eqnarray}
with  $F(R)\equiv df(R)/dR$ and $\mathrm{T}^{m}_{\mu\nu}$ the
standard matter stress-energy tensor derived from the matter
Lagrangian $\mathfrak{L}_{m}$ in the action (1). The vacuum
equations of motion, i.e. in the absence of matter, are given by,
\begin{eqnarray}
F(R){R}_{\mu\nu}-\frac{1}{2}f(R)g_{\mu\nu}\nabla_{\mu}\nabla_{\nu}F(R)+g_{\mu\nu}\square
F(R)=0
\end{eqnarray}
Contraction of the above field equations gives the following
relation between $f(R)$ and its derivative $F(R)$
\begin{eqnarray}
F(R)R-2f(R)+\square F(R)=0
\end{eqnarray}
which will be employed later both to simplify the field equations
and to find the general form of the $f(R)$ function.
\section{Cylindrically symmetric vacuum solutions}
Interested in the static cylindrically symmetric solutions of the
vacuum field equations (4), we start with the general form of such a
metric in the cylindrical  coordinates $(t,r,\varphi,z)$ given by
\cite{6};
\begin{eqnarray}\label{metr}
ds^{2}=-A(r)dt^{2}+dr^{2}+B(r)d\varphi^{2}+C(r)dz^{2}
\end{eqnarray}
The corresponding scalar curvature is
\begin{eqnarray}
R=\sum^{3}_{(i\neq
j)=1}(\frac{A''_{i}}{A_{i}}+\frac{1}{2}\frac{A'_{i}}{A_{i}}.\frac{A'_{j}}{A_{j}}-\frac{1}{2}(\frac{A'_{i}}{A_{i}})^{2})
\end{eqnarray}
in which $' \equiv\frac{d}{dr}$ and $A_{i}(r)=A(r),B(r),C(r)$ for
$i=1...3$  are the metric functions. Using equation $(5)$, the
modified Einstein equations become
\begin{eqnarray}
F
{R}_{\mu\nu}-\nabla_{\mu}\nabla_{\nu}F=\frac{1}{4}g_{\mu\nu}(FR-\square
F(R))
\end{eqnarray}
As in the spherical case, since the metric only depends on the
cylindrical radial coordinate $r$, one can view Eq. (8) as a set of
differential equations for functions $F(r)$, $u(r)$, $k(r)$ and
$w(r)$. In this case both sides are diagonal and hence we have four
equations. Differentiating Eq. (5) with respect to $r$  we have the
extra consistency relation for $F(r)$,
\begin{eqnarray}
R\acute{F}-\acute{R}F+3(\square{F})^{'}=0
\end{eqnarray}
Any solution of Eq. (8) must satisfy this relation in order to be
also a solution of the original modified Einstein's equations. From
Eq. (8) it is found that
\begin{eqnarray}
\frac{F
{R}_{\mu\mu}-\nabla_{\mu}\nabla_{\mu}F}{g_{\mu\mu}}=\frac{1}{4}(FR-\square
F(R))
\end{eqnarray}
In other words the combination $A_{\mu} \equiv \frac{F
{R}_{\mu\mu}-\nabla_{\mu}\nabla_{\mu}F}{g_{\mu\mu}}$ (with fixed
indices) is independent  of the index $\mu$ and therefore
$A_\mu=A_\nu$ for all $\mu,\nu$. This allows us to write the
following independent field equations;
\begin{eqnarray}
F''+\frac{1}{4}[\frac{A'}{A}(\frac{B'}{B}+\frac{C'}{C})-2(\frac{B''}{B}+\frac{C''}{C})+((\frac{B'}{B})^{2}
+(\frac{C'}{C})^{2})]F=0\\
 F'+\frac{1}{2}\frac{B'}{B}[2(\frac{A''}{A}-\frac{B''}{B})+((\frac{B'}{B})^{2}-(\frac{A'}{A})^{2})
 +\frac{C'}{C}(\frac{A'}{A}-\frac{B'}{B})]F=0\\
F'+\frac{1}{2}\frac{C'}{C}[2(\frac{A''}{A}-\frac{C''}{C})+
((\frac{C'}{C})^{2}-(\frac{A'}{A})^{2})+\frac{B'}{B}(\frac{A'}{A}-\frac{C'}{C})]F=0
\end{eqnarray}
corresponding to $A_t = A_r$, $A_t = A_\phi$ and $A_t = A_z$
respectively. Therefore, any set of functions $F(r)$, $A(r)$, $B(r)$
and $C (r)$ satisfying the above equations would be a solution of
the modified Einstein field equations (8). Obviously these equations
can not be solved without auxiliary conditions reducing the number
of the unknown functions. In the following section we discuss the
simple but important case of the solutions with constant curvature.
\subsection{A brief discussion about  cylindric constant curvature
solution  for couple realistic F(R) models}

One reason for investigating the constant curvature solution in a
typical modified modified f(R) gravity is that , in order that the
accelerating expansion in the present universe could be generated,
we must  consider that f(R) could be a small constant at present
universe \cite{20}.In this class of models, the universe starts from
the inflation driven by the effective cosmological constant
 at the early stage, where curvature is very large. As
curvature becomes smaller, the effective cosmological constant also
becomes smaller. After that the radiation/matter dominates. When the
density of the radiation and the matter becomes small and the
curvature goes to the value $R_{0}$  , there appears the small
effective cosmological constant. Hence, the current cosmic expansion
could start.In the epoch of the inflation, the curvature $R = A$
could be large , we may identify $f(\infty)$ as the cosmological
constant for the inflationary epoch and $f(R_{0})$ as that at the
present accelerating era \cite{20}.

\subsection{Constant Curvature solutions in Cylindrical symmetry  }
It is known that some of the vacuum constant curvature solutions in
$f(R)$ gravity are equivalent to  vacuum solutions in Einstein
theory with the same symmetry. For example it is shown in \cite{4}
that in the spherically symmetric case the corresponding $f(R)$
solutions include the  Schwarzschild-de-Sitter space for a specific
choice of one of the constants of integration. For cylindrical
symmetry, in Einstein gravity, static vacuum solutions were found
almost immediately after their spherical counter parts by
Levi-Civita \cite{6} but those with a cosmological constant have to
wait another 60 years to be found by Linet \cite{4} and
independently by Tian \cite{5} in a non-Weyl coordinate system.
Their solution reduces to that
 of a cosmic string in the limit $r \rightarrow 0$  \cite{7,8}.\\
Looking for cylindrically symmetric solutions in $f(R)$ gravity,
here we consider the simple but physically important case of static
constant curvature spacetimes. To do so, in the field equations
(11), (12) and (13) taking $R=constant$, we arrive at the following
set of equations:
\begin{eqnarray}
\frac{A'}{A}(\frac{B'}{B}+\frac{C'}{C})-2(\frac{B''}{B}+\frac{C''}{C})+((\frac{B'}{B})^{2}
+(\frac{C'}{C})^{2})=0\\
2(\frac{A''}{A}-\frac{B''}{B})+((\frac{B'}{B})^{2}-(\frac{A'}{A})^{2})
 +\frac{C'}{C}(\frac{A'}{A}-\frac{B'}{B})=0\\
2(\frac{A''}{A}-\frac{C''}{C})+
((\frac{C'}{C})^{2}-(\frac{A'}{A})^{2})+\frac{B'}{B}(\frac{A'}{A}-\frac{C'}{C})=0
\end{eqnarray}
These equations must be solved . One simple assumption is seeking
power law solutions for metric functions i.e. $A_{i}(r)\propto
r^{m_{i}}$ which leads to 3 simple algebraic equations for $m_{i}$.
which as we show in next section this solution and another
solutions. In the rest of this section we show that this solution
with another set of solutions correspond respectively to a zero and
non-zero values of Ricci scalar $R$.
\subsection*{Case (1) : solution with $R=0$}
From the  equations (14-16) one could obviously arrange a solution
of $A_{i}(r)\propto r^{m_{i}}$   in which  the constants $m_{i}$ are
given as follows;
\begin{eqnarray}
m_{1}=m_{3}=0,m_{2}=2\\
m_{1}=m_{2}=\frac{4}{3},m_{3}=-\frac{2}{3}
\end{eqnarray}
It can be shown that these are  Ricci flat solutions (i.e $R=0$ in
(7)) .Indeed in the second case which
$m_{1}=m_{2}=\frac{4}{3},m_{3}=-\frac{2}{3}$ the metric is written
as:
\begin{eqnarray}
\label{metric2}
ds^2=-A_{0}r^{4/3}dt^{2}+dr^{2}+B_{0}r^{4/3}d\varphi^{2}+C_{0}r^{2/3}dz^{2}
\end{eqnarray}
by applying the following  radial coordinate transformation
\begin{eqnarray}
r=\frac{4}{3}\rho^{3/4}
\end{eqnarray}
and choosing the constants
$A_{0}=B_{0}=(\frac{3}{4})^{4/3},C_{0}=(\frac{3}{4})^{2/3}$
 it transforms into the following historic  metric
\begin{eqnarray}\label{metric4}
ds^2 = -\rho
dt^2+\rho^{-1}[\rho^{1/2}(dz^2+d\rho^2)+\rho^2d\varphi^2]
\end{eqnarray}
it is Levi-Civita's static cylindrically symmetric solution which
normally is written in the following general form with the constant
$m=\frac{1}{2}$:
\begin{eqnarray}\label{metric5}
ds^2 =
-\rho^{2m}dt^2+\rho^{-2m}[\rho^{2m^2}(d\rho^2+dz^2)+\rho^2d\varphi^{2}]
\end{eqnarray}
 That it is obviously flat \cite{10}.
\subsection*{Case (2) : solution with $R = constant \neq 0$}
To find solutions of this type we introduce 3 new functions as   $
A=e^u,B=e^v,C=e^w$. In terms of these functions the equations
(11-13) become:
\begin{eqnarray}
u'(v'+w')-2(v''+w'')-(v'^2+w'^2)=0\\
2(u''-v'')-(v'^2-u'^2)+w'(u'-v')=0\\
2(u''-w'')+(u'^2-w'^2)+v'(u'-w')=0
\end{eqnarray}
Solving this equations is complicated. But in a special case $u=v$
equation (24) is satisfied identically and 2 remaining
equations(23,25) reduce to:
\begin{eqnarray}
u''-\frac{1}{2}u'w'+w''+\frac{1}{2}w'^2=0\\
u''-\frac{1}{2}u'w'-w''-\frac{1}{2}w'^2+u'^2=0
\end{eqnarray}
By adding  (26,27) we have:
\begin{eqnarray}
2u''+u'^2-u'w'=0
\end{eqnarray}
a straightforward integration gives:
\begin{eqnarray}
w=2\ln u'+u+c
\end{eqnarray}
in which  $c$ is a constant . After substituting this solution in
 (26) we have:
\begin{eqnarray}
3u''+2\frac{u^{'''}}{u'}=0
\end{eqnarray}
 this is a simple 2'nd order ODE w.r.t $u'$.Taking  $ p=u'$ and
 rewriting the derivatives with respect to $u$  we have:
\begin{eqnarray}
\frac{2}{3}p\frac{dp}{du}=c-\frac{1}{2}p^2
\end{eqnarray}
which has the following solution:
\begin{eqnarray}
p=\pm\sqrt{2c+Ae^{-3u/2}}
\end{eqnarray}
If we choose only positive sign we can write an explicit form for
$u,v,w$:
\begin{eqnarray}
u=v=\frac{2}{3}\ln[\frac{A}{2c}\cos^{2}(\frac{3}{2}\sqrt{\frac{c}{2}}.r)]\\
w=2\ln[\sqrt{2c}.\tan(\frac{3}{4}\sqrt{2c}.r)]+
\frac{2}{3}\ln[\frac{A}{2c}\cos^{2}(\frac{3}{4}\sqrt{2c}.r)]+c_{2}
\end{eqnarray}
Without loss of generality  we can set $c_{2}=0$ or absorb it in the
$ z$ coordinate. If we calculate Ricci scalar (7) with these metric
functions we obtain $ R=6c$.So by comparing this with the solutions
of the Einstein field equation  in the presence of the cosmological
 constant i.e,  $R=4\Lambda$,we have:
\begin{eqnarray}
c = \frac{2}{3} \Lambda
\end{eqnarray}
Finally we can write the general form of metric (6) as:
\begin{eqnarray}
ds^2=-[\frac{A}{2c}\cos^{2}(\frac{3}{2}\sqrt{\frac{c}{2}}.r)]^{2/3}dt^2+dr^2+
[\frac{A}{2c}\cos^{2}(\frac{3}{2}\sqrt{\frac{c}{2}}.r)]^{2/3}d\varphi^2
+\\\nonumber[2A^2c \sin^2(3/4\sqrt{2c} r)\cos^{4/3}(3/4\sqrt{2c}
r)]dz^2
\end{eqnarray}
Now we introduce new parameter $\alpha=(\frac{A}{2c})^{2/3}$. Our
metric solution converts to:
\begin{eqnarray}
ds^2=\alpha
\cos^{4/3}(\sqrt{\frac{3\Lambda}{4}}r)(d\varphi^2-dt^2)+dr^2+(\frac{4}{3}\Lambda\alpha)^3\sin^2(\sqrt{\frac{3\Lambda}{4}}r)
\cos^{4/3}(\sqrt{\frac{3\Lambda}{4}}r)dz^2
\end{eqnarray}
 from the general form of Tian metric we have:
\begin{eqnarray}
ds^2=-[\tan \beta(r+\hat{r})]^{\gamma_{1}}[\sin
2\beta(r+\hat{r})]^{2/3}dt^2+[\tan
\beta(r+\hat{r})]^{\gamma_{2}}[\sin
2\beta(r+\hat{r})]^{2/3}dz^2\\\nonumber+[\tan
\beta(r+\hat{r})]^{\gamma_{3}}[\sin
2\beta(r+\hat{r})]^{2/3}d\varphi^2+dr^2
\end{eqnarray}
 Where  $\hat{r}$ is an arbitrary real number, $\beta=\sqrt{\frac{3\Lambda}{4}}$ and
 $\gamma_{1},\gamma_{2}$ $\gamma_{3}$ are real numbers in the
 interval  $[-\frac{4}{3},\frac{4}{3}]$,   satisfying the algebraic
 equations
\begin{eqnarray}
\gamma_{1}+\gamma_{2}+\gamma_{3}=0,\gamma_{1}\gamma_{2}+\gamma_{2}\gamma_{3}+\gamma_{1}\gamma_{3}=-\frac{4}{3}
\end{eqnarray}
Easily we observe that our solution is a special sub family of Tian
solution with parameters
$\gamma_{1}=\gamma_{3}=-\frac{2}{3},\gamma_{2}=\frac{4}{3},\hat{r}=0$
, $\alpha=\frac{3\sqrt{2}}{4\Lambda}$ . Also  a change of
coordinates (it is better to called them a congruity of non radial
coordinates )is needed  as:
\begin{eqnarray}
\varphi\rightarrow
(\frac{\sqrt{2}\Lambda}{3})^{1/2}\varphi,t\rightarrow
(\frac{\sqrt{2}\Lambda}{3})^{1/2}t
\end{eqnarray}
Then we prove that in $f(R)$  gravity the constant curvature
solution in cylindrically symmetric cases is only one member of the
most generalized Tian family in GR, thus we have different kinds of
cosmological constant in this scenario and not a fixed one as in
GR.
\section{Mathematical behavior of Field equations as an autonomous dynamical system}
In this section our goal is description stability of system of ODE
(23-25) in modern language of dynamical systems \cite{14}. A non
linear system of differential equations   $\acute{X}=f_{i}(X)
,X=(X_{1},X_{2},....,X_{n}) ,\acute{X}=\frac{d X}{d t}$ is called
autonomous . A point $ P=(X^{0}_{1},X^{0}_{2},....,X^{0}_{n}) $ is
called a critical point if that be a solution for the system of
algebraic equations:
\begin{eqnarray}
f_{i}(X)=0, i=1,2,...,n
\end{eqnarray}
any  autonomous system  can be spotted near critical point as a
linear system. If we construct a matrix $A$ with components
$A_{ij}=(\frac{\partial f_{i}(X)}{\partial X_{j}})|_{P}$ , and if
the linear perturbation of system from critical point is denoted by
$\Psi=(\Psi_{1},\Psi_{2},....,\Psi_{n}) $, then the solution to
linearized system $\acute{\Psi}=A\Psi$ determines stability of the
original non linear model. If as $t\rightarrow\infty$ all
perturbation functions $\Psi_{i}\rightarrow 0$ then we say that the
system is stabilized. For our purpose if in system (23-25) we
introduce a new set of variables $\xi=u',\eta=v',\theta=w'$ and if
we denote all of them in form $\xi_{h},i=1,2,3$ we can write this
equations in form :
\begin{eqnarray}
\frac{d\xi_{i}}{dr}=\frac{1}{2}(\xi_{j}\xi_{k}-\xi_{i}^2),j\neq
k,i,j,k=1,2,3
\end{eqnarray}
it's related matrix $A$ is:
\begin{eqnarray}
A_{ij}=\frac{1}{2}\epsilon_{ijk}\xi_{k}-\xi_{i}\delta_{ij}
\end{eqnarray}
 Where in it $\epsilon_{ijk}$ is a Levi-Civita altering tensor and
 $\delta_{ij}$ is Kronecher delta function.
 The critical point of system is located at $ P=(\xi,\eta,\theta)=c(1,1,1)$ where in
 it  $c$ is an arbitrary constant. Thus it's value at critical point
  can be written explicitly by:
\begin{eqnarray}
A_{i=j}=-c,A_{i\neq j}=\frac{1}{2}c
\end{eqnarray}
the eigenvalues of matrix $A$ are:
\begin{eqnarray}
\lambda_{1}=0,\lambda_{2,3}=-\frac{3}{2}
\end{eqnarray}
Thus the general solution for system (42) in linearized aproximation
is :
\begin{eqnarray}
\Psi=\xi_{0}+(\eta_{0}+\theta_{0})e^{-\frac{3}{2}r}
\end{eqnarray}
 For stability we observe that only if $\xi_{0}=0$ the non linear
 system (42) in linear approximation is stable and consequently the
 non linear system begins stable.
\section{Physical results and  Conclusions}
In recent decades there has been more interests in Cosmic strings as
an Cosmological predictions of String theory. We know that according
to the gauge theories with spontaneous symmetry breaking the
universe has a number of phase transitions since the big bang. In
general, cosmic strings are topological structures which is produced
meanwhile in these phase transitions. Later Grfinkle \cite{15} has
treated the  string as a self-interacting massless minimally coupled
scalar field coupled to a U(1) gauge field and has shown that there
exists a class of static , cylindrically symmetric solutions to the
field equations.
 Now following  the Tian \cite{5} we prove that our metric (37) is applicable to the exterior of a
 string. This is the main physics which is hidden behind
 mathematical equations and we elucidate it in the following
 paragraphes.
Consider the exterior metric for a string. One class of exteriors
has been  discussed by Tian  is a solution of  equations
 in a vaccum or by imposing the boost symmetry. The boost
symmetry means that in the general Tian metric we must have
\begin{eqnarray}\nonumber
\gamma_{1}=\gamma_{2}=\pm\frac{2}{3} , \\ \nonumber
\gamma_{3}=\pm\frac{4}{3}
\end{eqnarray}
The result metric is:
\begin{eqnarray}\nonumber
ds^2=[a\cos\beta(r+\hat{r})]^{4/3}(-dt^2+dz^2)+\frac{d^2a^{-2/3}}{\beta^2}[\sin\beta(r+\hat{r})]^2\times
[\cos\beta(r+\hat{r})]^{-2/3}d\varphi^2+dr^2
\end{eqnarray}
If we apply the complex transformations $t\rightarrow
-i\varphi,z\rightarrow it,\varphi\rightarrow z$, this form
accordance with the our metric (37) by choosing parameters
$\hat{r}=0$ ,
$\alpha=\frac{3\sqrt{2}}{4\Lambda},a=\alpha^{3/4},d=\beta
a^{1/3}$.This proofs that our metric (37) is applicable to the
exterior of a string.\\
 Again the motivation for considering material systems with cylindrical symmetry in asymptotically AdS
spacetimes comes also from the 'classical general relativity'.
 it is well known that with  $\Lambda=0$ , the asymptotes of cylindrically symmetric static spacetimes representing
infinite sources is very different from that of spatially bounded
static sources. The advantage of the studied spacetimes over the
vaccum LC case is that for  $\Lambda<0$, far away from the symmetry
axis, $r = 0$ , they approach the anti de Sitter solution (AdS)
unlike LC that does not approach the Minkowski spacetime. Lemos
\cite{16} found an interesting class of cylindrically symmetric 4D
solutions with  $\Lambda<0$ called black strings. These spacetimes
describe the fields of charged, rotating strings that generally
feature singularities and horizons and are asymptotically anti de
Sitter far away from the axis of symmetry. Static plane-symmetric
perfect fluids occur also as subcases of the static
cylindrically-symmetric solutions, There are several physically
interesting classes of solutions admitting an Abelian group $G(2)$
on non-null orbits; because of their importance and the large amount
of relevant material, they will be divided into separate parts. The
stationary axisymmetric fields have timelike group orbits $T^2$. The
classes with spacelike group orbits $S^2$ Contains time-dependent
cylindrically-symmetric fields and their stationary subclasses,
colliding plane waves with their typical dependence on the retarded
advanced time, and inhomogeneous perfect fluid solutions. Known
rigidly rotating axisymmetric perfect fluid solutions with more than
two Killing vectors belong to the locally rotationally-symmetric
space-times or to the homogeneous space-times or they are
cylindrically-symmetric .
 The stationary cylindrically-symmetric dust solutions are contained in
the general stationary axisymmetric dust solution (the Winicour
solution)\cite{17}.
  About generalization of our cylindrically symmetric  result for other modified
 gravities especially for $f(G)$ and  $f(G,R)$ one could obtain  new families of
 physically tolerable solution.Former stationary cylindrically symmetric solutions to the five-dimensional
  Einstein and Einstein - Gauss - Bonnet equations has been investigated and two exact solutions which qualify as cosmic strings,
  one corresponding to an electrically charged cosmic string,
   the other to an extended superconducting cosmic string surrounding a charged core were found\cite{18}.
    Thus naturally we expected that in any $f(G)$ model of gravity
   by imposing the cylindrical symmetry one could be able to recover
   this rather old family of exact solutions. Note that in $f(G)$-gravity, there are no problems
    with the Newton law, instabilities and the anti-gravity regime
    \cite{21}.The analogue of constant curvature in $f(G)$-gravity is $G_{0}$  corresponds to
    the present value of the Gauss-Bonnet invariant. Technically  when $G = G_{0}$ and $G = G_{+\infty}$,
    $f(G)$ becomes almost constant and can be regarded as the effective
   cosmological constant similar to $f(R)$ models.
  The starting point for doing it is following from
   which  has been done  by   Nojiri and
    Odintsov for  spatially-flat FRW universe metrics in the context of
   modified Gauss-Bonnet $f(G)$ and  $f(G,R)$ gravity\cite{19}. As an landscape it seems
   that for a large number of choices of the function
    $f(G)$, like that has been discussed in \cite{19,21} , the field equations has a
    non-trivial real solution  (in which non of metric functions vanishes)
    and asymptotically converge to  a  static patch of deSitter universe.
    We remind it to the another work \cite{22}.

\section{ Acknowledgment}
D.Momeni thanks S. Nojiri  and S. D .Odintsov for useful
 comments. Finally, the editor of IJMPD , Jorge Pullin and the anonymous referees made excellent observations and
suggestions which resulted in substantial improvements of the
presentation and the results.


\end{document}